\documentclass[conference]{IEEEtran}
\IEEEoverridecommandlockouts
\usepackage[comma, sort&compress,square,numbers]{natbib}
\usepackage{amsmath,amssymb,amsfonts}
\usepackage{algorithmic}
\usepackage{graphicx}
\usepackage{textcomp}
\usepackage{xcolor}
\usepackage{hyperref}
\usepackage{threeparttable}
\usepackage{makecell}
\usepackage{colortbl}
\usepackage{booktabs}
\usepackage{graphicx} 
\usepackage{multirow} 
\usepackage{rotating} 
\usepackage{svg}
\usepackage{subfigure}
\usepackage{authblk}
\usepackage{soul}
\usepackage{flushend}

\def\BibTeX{{\rm B\kern-.05em{\sc i\kern-.025em b}\kern-.08em
    T\kern-.1667em\lower.7ex\hbox{E}\kern-.125emX}}
\begin{document}

\title{Benchmarking Conventional and Learned Video Codecs with a Low-Delay Configuration\\
}

\author[1]{Siyue Teng}
\author[1]{Yuxuan Jiang}
\author[1]{Ge Gao}
\author[1]{Fan Zhang}
\author[2]{Thomas Davis}
\author[2]{Zoe Liu}
\author[1]{David Bull}
\affil[1]{\textit{Visual Information Laboratory, University of Bristol, Bristol, BS1 5DD, United Kingdom}}
\affil[1]{\textit {\{Siyue.Teng, Yuxuan.Jiang, Ge1.Gao, Fan.Zhang, Dave.Bull\}@bristol.ac.uk}}
\affil[2]{\textit{Visionular Inc., Los Altos, CA 94022 USA}}
\affil[2]{\textit {\{thomas, zoeliu\}@visionular.com}}


\maketitle

\begin{abstract}

Recent advances in video compression have seen significant coding performance improvements with the development of new standards and learning-based video codecs. However, most of these works focus on application scenarios that allow a certain amount of system delay (e.g., Random Access mode in MPEG codecs), which is not always acceptable for live delivery. This paper conducts a comparative study of state-of-the-art conventional and learned video coding methods based on a low delay configuration. Specifically, this study includes two MPEG standard codecs (H.266/VVC VTM and JVET ECM), two AOM codecs (AV1 libaom and AVM), and two recent neural video coding models (DCVC-DC and DCVC-FM). To allow a fair and meaningful comparison, the evaluation was performed on test sequences defined in the AOM and MPEG common test conditions in the YCbCr 4:2:0 color space. The evaluation results show that the JVET ECM codecs offer the best overall coding performance among all codecs tested, with a 16.1\% (based on PSNR) average BD-rate saving over AOM AVM, and 11.0\% over DCVC-FM. We also observed inconsistent performance with the learned video codecs, DCVC-DC and DCVC-FM, for test content with large background motions. 
\end{abstract}


\section{Introduction}

In recent years, video content has been the dominant contributor to global data traffic. This has been amplified by the latest advances in live application scenarios such as online streaming, sports broadcasting, real-time surveillance, and video conferencing. Live transmission imposes delivery challenges, particularly for video compression~\cite{bull2021intelligent}, because video encoding and transmission typically target minimum system latency. 

The past two decades have seen the development of multiple generations of video coding standards. While MPEG H.264/AVC~\cite{h264AVC}, initially introduced twenty years ago, is still widely used in many applications, two further coding standards, H.265/HEVC~\cite{h265HEVC} and H.266/VVC~\cite{h266VVC}, have been released with significantly improved coding performance. In parallel, the Alliance for Open Media (AOM), formed in 2015, has delivered its first video coding standard, AOMedia Video 1 (AV1). More recently, both MPEG and AOM have initiated the exploration of new video coding algorithms beyond their latest standards, with working codecs ECM (Enhanced Compression Model)~\cite{ecm_github} and AVM (AOM Video Model)~\cite{avm_github}, respectively.

Alongside these standard video codecs, there has been a rapid emergence of neural video coding methods, inspired by breakthroughs in deep learning, which are now competing with conventional coding algorithms. Notable work includes early attempts such as DVC~\cite{dvc}, DVCPro~\cite{dvcpro} and FVC~\cite{fvc}, followed by further advancements, such as transformer-based models~\cite{vct,mimt,zhu2022transformer}, the normalizing flow-based CANF series~\cite{canfvc,chen2023canf}, implicit neural representation based codecs \cite{kwan2023hinerv,chen2021nerv} and conditional coding based DCVC methods~\cite{li2021deep, sheng2022temporal, li2022hybrid, li2023neural, li2024neural}.

\begin{figure}[!t]
    \centering
    \includegraphics[width=1\linewidth]{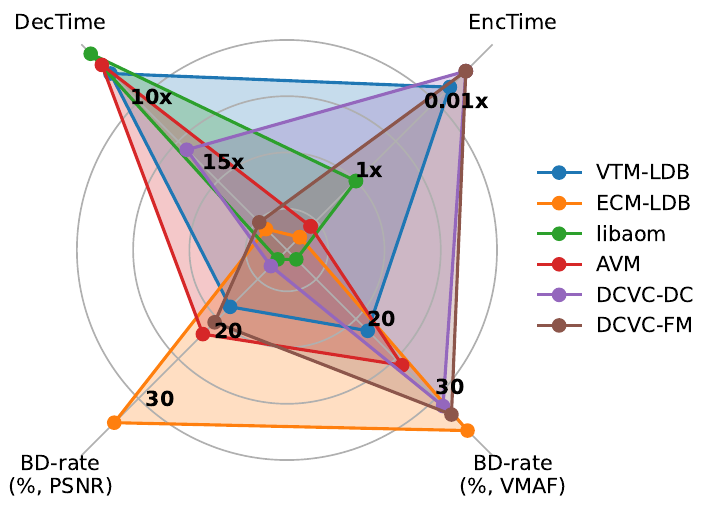}
    \caption{Radar chart plot for six tested codecs in terms of BD-rate (PSNR), BD-rate (VMAF), relative encoding complexity and decoding complexity (all against libaom). Larger shaded area indicates better performance-complexity trade off for the corresponding codec. }
    \label{fig:radar-chart}
\end{figure}

\begin{table*}[!t]
  \centering
  \caption{Coding configurations employed in the benchmark experiment.}
    \resizebox{0.99\textwidth}{!}{\begin{tabular}{l|l|l|l}
    \toprule
    \multicolumn{1}{l|}{Codec} & \multicolumn{1}{l|}{Version} & \multicolumn{1}{l|}{QP} & Coding Parameters \\
    \midrule
    VTM   & 23.1     & \makecell[l]{22 27 32 37} & \multicolumn{1}{l}{\multirow{2}[2]{*}{Default main10 profile with Low-Delay P (LDP) and Low-Delay B (LDB) configuration, Level=6.2.}} \\
    \cmidrule{1-3}
    ECM   & 12.0     & \makecell[l]{22 27 32 37} &  \\
    \midrule
    libaom   & \makecell[l]{3.8.1-324-\\g732e092}     & \makecell[l]{27 35 46 55 63} & \multicolumn{1}{l}{\multirow{2}{*}{\makecell[l]{-{}-cpu-used=0 -{}-passes=1 -{}-lag-in-frames=0 -{}-min-gf-interval=16 -{}-max-gf-interval=16 -{}-gf-min-pyr-height=4\\ -{}-gf-max-pyr-height=4 -{}-limit=\$frame\_num  -{}-kf-min-dist=9999 -{}-kf-max-dist=9999 -{}-use-fixed-qp-offsets=1\\ -{}-deltaq-mode=0 -{}-enable-tpl-model=0, -{}-end-usage=q \textcolor{orange}{-{}-cq-level=\${qp}} \textcolor{orange}{(for libaom)} \\ \textcolor{cyan}{-{}-qp=\${qp} -{}-subgop-config-str=ld} \textcolor{cyan}{(for AVM)} -{}-enable-keyframe-filtering=0  -{}-obu -{}-tile-columns=0 -{}-threads=1}}} \\
    \cmidrule{1-3}
    AVM   & 2.0.0     & \makecell[l]{110 135 160 185\\ 210 235} &  \\
    \midrule
    DCVC-DC & \cellcolor[HTML]{EFEFEF}   & \makecell[l]{0 18 36 63} & \makecell[l]{-{}-i\_frame\_model\_path ./checkpoints/cvpr2023\_image\_yuv420\_psnr.pth.tar \\ -{}-p\_frame\_model\_path ./checkpoints/cvpr2023\_video\_yuv420\_psnr.pth.tar -{}-rate\_num 8 \\ -{}-test\_config ./\textless config\textgreater .json -{}-yuv420 1 -{}-cuda 1 -{}-worker 1 -{}-write\_stream 1 -{}-output\_path \textless output\textgreater .json \\ -{}-force\_intra\_period 9999 -{}-save\_decoded\_frame 1 -{}-calc\_ssim 1 -{}-verbose 1} \\
    \midrule
    DCVC-FM   & \cellcolor[HTML]{EFEFEF}     & \makecell[l]{27 45 54 63} & \makecell[l]{-{}-model\_path\_i ./checkpoints/cvpr2024\_image.pth.tar \\-{}-model\_path\_p ./checkpoints/cvpr2024\_video.pth.tar -{}-rate\_num 8 -{}-test\_config ./\textless config\textgreater .json\\ -{}-cuda 1 -{}-worker 1 -{}-write\_stream 1 -{}-output\_path \textless output\textgreater .json\\ -{}-force\_intra\_period 9999 -{}-save\_decoded\_frame 1 -{}-calc\_ssim 1 -{}-verbose 1 -{}-verbose\_json 1} \\
    \bottomrule
    \end{tabular}}
  \label{tab: codecConfig}
\end{table*}%
To benchmark the coding performance of the above-mentioned video codecs, existing works~\cite{2015LD, 2020LD5codec, katsenou2019subjective, katsenou2022bvi} employ different coding configurations and datasets, leading to inconsistent results and conclusions. It is also noted that most of these studies~\cite{2024HRRA, 2021RA3codec, 2019RA2codec} allow relatively high system delays (e.g., following the Random Access mode~\cite{aom_ctc_v3, vtm_ctc} with MPEG/AOM test models), while evaluation for low-latency scenarios has not been extensively performed.

In this paper, specifically targeting low delay applications, we conducted a comprehensive evaluation of six standard and neural video codecs including VVC VTM, JVET ECM, AOM AV1 libaom, AOM AVM, DCVC-DC, and DCVC-FM. The benchmark experiment is based on the common test conditions defined by the MPEG JVET and AOM standard bodies~\cite{vtm_ctc,aom_ctc_v3}, which were designed to align with real video streaming applications. To ensure a fair and meaningful comparison, all encodings were implemented in the YCbCr 4:2:0 color space. Coding performance is measured using the BD (Bj{\o}ntegaard Delta)-rate metric~\cite{bdrate} based on two objective quality metrics, PSNR and VMAF. The results show that, as illustrated in \autoref{fig:radar-chart}, ECM offers the best overall coding performance among all codecs tested, although it is also associated with a high encoding and decoding complexity. None of these tested codecs offer a perfect trade off between complexity and coding efficiency - this emphasizes the need to develop new video coding methods for low latency constraint with low computational (in particular decoding) complexity and high coding performance.   

The remainder of the paper is organized as follows. \autoref{exp setup} details the experimental configuration for evaluating the selected codecs. Section III summarizes and analyzes the results in terms of coding performance and computational complexity. Finally, Section IV concludes the paper by highlighting the key findings and implying potential future works.

\section{Experimental setup}\label{exp setup}

This section specifies the codec selection and configurations, test sequences, evaluation metrics, and experimental hardware.

\subsection{Codec Selection and Configuration}

In this experiment, six video codecs were selected, including H.266/VVC VTM~\cite{vtm_github}, JVET ECM~\cite{ecm_github}, AOM AV1 libaom~\cite{av1_github}, AOM AVM~\cite{avm_github}, DCVC-DC~\cite{li2023neural} and DCVC-FM~\cite{li2024neural}. Among these, VTM is the test model of the latest MPEG video coding standard, H.266/Versitle Video Coding (VVC), while ECM is Enhanced Compression Model which is being developed by MPEG JVET. These two represent the state-of-the-art from MPEG standards. On the other hand, libaom is the reference encoder of the AV1 standard, and AVM is the AOM video model under development for the next AOM video coding standard. These offer the best coding performance among the AOM codecs. 

Regarding neural video codecs, we have selected two codecs, DCVC-DC~\cite{li2023neural} and DCVC-FM~\cite{li2024neural}, which have been reported to exhibit superior performance under low delay conditions. More importantly, they support compression in the YCbCr 4:2:0 color space, which enables a fair comparison with conventional video coding methods.

For all six video codecs benchmarked, we employ their latest versions and perform encoding based on the Low Delay mode defined by the MPEG JVET and AOM common test conditions (CTC)~\cite{aom_ctc_v3, vtm_ctc}, which requires the encoding order in a sequence to be the same as the temporal order (no B frames are allowed).  Detailed information in terms of codec versions, quantization parameters (QP), and coding parameters is summarized in \autoref{tab: codecConfig}. For two MPEG codecs, VTM and ECM, there are two low delay modes available, denoted as LDP and LDB. We used both in our experiment. For two neural video codecs, we used their pre-trained models released associated with their original literature, without performing further fine-tuning.

\subsection{Test Sequences}

In this experiment, we selected 53 test sequences defined in the JVET and AOM CTCs~\cite{aom_ctc_v3, vtm_ctc} for the Low Delay configuration due to their diversity and coverage of the content. Specifically, there are 13 video clips from JVET (Class B, C, and D) and 40 from AOM (Classes A2, A3, A4, and A5), covering resolutions from 480p to 1080p, with 8 or 10 bit depth. It is noted that we did not include higher resolution content, e.g., UHD videos, in this test, due to resource and computational complexity constraints with the neural video codecs.

\subsection{Evaluation Metrics}

\begin{figure*}[t]
    \centering
    \begin{minipage}[c]{0.68\textwidth}
        \centering
          {\footnotesize TABLE II \\\textsc{Coding performance and coding complexity results. Here the BD-rate value in each cell is calculated using the codec in this row as the anchor.}}
          \vspace{6pt}
          
        \resizebox{\columnwidth}{!}{
            \begin{tabular}{r|r|r|r|r|r|r|r|c}
            \toprule
            \multicolumn{8}{c|}{PSNR} & \multicolumn{1}{c}{Encoding time}\\
            \cmidrule{1-8}
        BD-rate   & \multicolumn{1}{c|}{DCVC-DC} & \multicolumn{1}{c|}{VTM-LDP} & \multicolumn{1}{c|}{VTM-LDB} & \multicolumn{1}{c|}{DCVC-FM} & \multicolumn{1}{c|}{AVM} & \multicolumn{1}{c|}{ECM-LDP} & \multicolumn{1}{c|}{ECM-LDB} & (HD only) \\
        \midrule
        libaom      & -11.2\% &-13.4\% & -19.2\% & -20.8\% & -21.4\% & -29.0\% & -33.9\% & 1.000$\times$\\
        \midrule  
        DCVC-DC  & \cellcolor[HTML]{EFEFEF} & 5.7\% & -0.4\% & -8.3\% & -2.9\% & -13.5\% & -18.8\% & 0.008$\times$\\
        \midrule  
        VTM-LDP  & \cellcolor[HTML]{EFEFEF} & \cellcolor[HTML]{EFEFEF} & -6.3\% & -7.2\% & -8.7\% & -17.8\% & -23.4\% & 1.032$\times$ \\
        \midrule  
        VTM-LDB  & \cellcolor[HTML]{EFEFEF} & \cellcolor[HTML]{EFEFEF} & \cellcolor[HTML]{EFEFEF} & -0.7\% & -2.4\% & -12.2\% & -18.3\% & 1.502$\times$\\
        \midrule  
        DCVC-FM  & \cellcolor[HTML]{EFEFEF} & \cellcolor[HTML]{EFEFEF} & \cellcolor[HTML]{EFEFEF} & \cellcolor[HTML]{EFEFEF}  & 5.8\% & -5.0\% & -11.0\% & 0.012$\times$\\
        \midrule  
        AVM      & \cellcolor[HTML]{EFEFEF} & \cellcolor[HTML]{EFEFEF} & \cellcolor[HTML]{EFEFEF} & \cellcolor[HTML]{EFEFEF} & \cellcolor[HTML]{EFEFEF} & -9.5\% & -16.1\% & 12.301$\times$\\
        \midrule  
        ECM-LDP  & \cellcolor[HTML]{EFEFEF} & \cellcolor[HTML]{EFEFEF} & \cellcolor[HTML]{EFEFEF} & \cellcolor[HTML]{EFEFEF} & \cellcolor[HTML]{EFEFEF} & \cellcolor[HTML]{EFEFEF} & -7.1\% & 10.704$\times$\\
        \midrule           
        ECM-LDB & \cellcolor[HTML]{EFEFEF} & \cellcolor[HTML]{EFEFEF} & \cellcolor[HTML]{EFEFEF} & \cellcolor[HTML]{EFEFEF} & \cellcolor[HTML]{EFEFEF} & \cellcolor[HTML]{EFEFEF} & \cellcolor[HTML]{EFEFEF} & 16.106$\times$\\
        
            \midrule
              \midrule
            \multicolumn{8}{c|}{VMAF} & \multicolumn{1}{c}{Decoding time}\\
            \cmidrule{1-8}
        BD-rate   & \multicolumn{1}{c|}{VTM-LDP} & \multicolumn{1}{c|}{VTM-LDB} & \multicolumn{1}{c|}{AVM} & \multicolumn{1}{c|}{ECM-LDP} & \multicolumn{1}{c|}{DCVC-DC}  & \multicolumn{1}{c|}{DCVC-FM} & \multicolumn{1}{c|}{ECM-LDB} & (HD only)\\
        \midrule
        libaom   & -18.2\% & -22.5\% & -25.6\% & -30.4\% & -31.1\% & -33.1\% & -34.2\% & 1.000$\times$ \\
        \midrule           
        VTM-LDP  & \cellcolor[HTML]{EFEFEF} & -4.6\% & -8.1\% & -14.5\% & -15.8\% & -18.1\% & -18.9\% & 2.723$\times$\\
        \midrule          
        VTM-LDB  & \cellcolor[HTML]{EFEFEF} & \cellcolor[HTML]{EFEFEF} & -3.4\% & -10.2\% & -10.6\% & -13.8\% & -15.0\% & 2.675$\times$\\
        \midrule         
        AVM    & \cellcolor[HTML]{EFEFEF} & \cellcolor[HTML]{EFEFEF} & \cellcolor[HTML]{EFEFEF} & -6.6\% & -8.4\% & -9.8\% & -11.8\% & 2.246$\times$\\
        \midrule          
        ECM-LDP  & \cellcolor[HTML]{EFEFEF} & \cellcolor[HTML]{EFEFEF} & \cellcolor[HTML]{EFEFEF} & \cellcolor[HTML]{EFEFEF} & -0.9\% & -3.4\% & -5.2\% & 21.397$\times$\\
        \midrule           
        DCVC-DC  & \cellcolor[HTML]{EFEFEF} & \cellcolor[HTML]{EFEFEF} & \cellcolor[HTML]{EFEFEF} & \cellcolor[HTML]{EFEFEF} & \cellcolor[HTML]{EFEFEF} & 0.0\% & 0.8\% & 13.180$\times$\\    
        \midrule           
        DCVC-FM  & \cellcolor[HTML]{EFEFEF} & \cellcolor[HTML]{EFEFEF} & \cellcolor[HTML]{EFEFEF} & \cellcolor[HTML]{EFEFEF} & \cellcolor[HTML]{EFEFEF} & \cellcolor[HTML]{EFEFEF} & 5.5\% & 20.735$\times$\\ 
        \midrule          
        ECM-LDB & \cellcolor[HTML]{EFEFEF} & \cellcolor[HTML]{EFEFEF} & \cellcolor[HTML]{EFEFEF} & \cellcolor[HTML]{EFEFEF} & \cellcolor[HTML]{EFEFEF} & \cellcolor[HTML]{EFEFEF} & \cellcolor[HTML]{EFEFEF} & 21.725$\times$\\
    \bottomrule
    \end{tabular}}
    
      \end{minipage}
    \hfill
    \begin{minipage}[c]{0.30\textwidth}
        \centering    
            \includegraphics[width=0.9\textwidth]{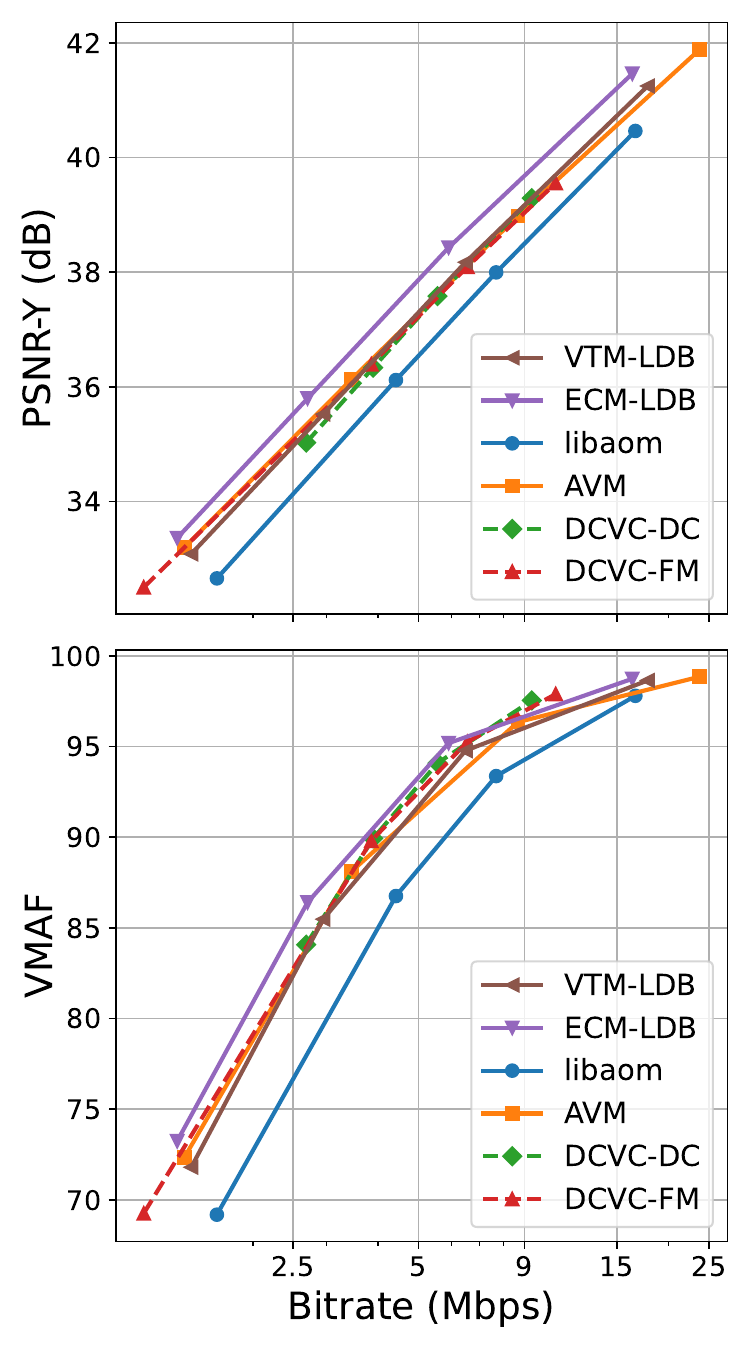}
        \vspace{-0.35cm}
        \caption{Rate-ristortion curve of HD sequences.}
        \label{Fig: rd-curve}
    \end{minipage}
\end{figure*}

The coding performance of tested codecs is measured based on their rate quality performance. Specifically, the video quality of each compressed video is calculated using two quality metrics, PSNR (Peak Signal-to-Noise Ratio, luma component only) and VMAF (Video Multimethod Assessment Fusion)~\cite{vmaf}. PSNR is the most widely used quality metric, which is recommended in JVET and AOM CTCs and is also widely used for evaluating the performance of neural video codecs. VMAF is employed in the AOM CTC, offering a more accurate prediction of perceived video quality by combining several existing quality metrics and a video feature using a learnable regressor. Based on the bitrate of the compressed bitstreams and the video quality indices, the Bj{\o}ntegaard Delta measurement~\cite{bdrate} is calculated for each sequence to compare the performance difference between two codecs.

Alongside coding performance, we also measured the average encoding and decoding runtimes for all tested codecs to compare their computational complexity. All the encodings and decodings for six benchmarked codecs were executed on a PC with an Intel i7-12700 CPU, 64GB RAM, and a NVIDIA 3090 GPU.

\vspace{-0.1cm}

\section{Results and Discussion}

This section provides the experimental results for the six tested video codecs in terms of their compression performance and computational complexity. We also observe and analyze specific scenarios where neural video codecs perform inconsistently.

\subsection{Overall Coding Performance}
\label{coding efficiency}

TABLE II summarizes the compression performance of all codecs tested; each cell presents the average BD-rate result (among all test sequences) of the codec in a column (test) against that in the row (anchor). It can be observed that, based on PSNR, ECM offers the best overall performance, with its LDB mode achieving a 18.3\% BD-rate saving over VTM (LDB), 11.0\% over DCVC-FM, 18.8\% over DCVC-DC and 16.1\% over AVM. The latter four codecs perform similarly, with minimal BD-rate values between each other \footnote{Inconsistent rankings are observed in TABLE II when the average BD-rate values are small, for example, when libaom is the anchor, AVM is slightly better than DCVC-FM, while the direct comparison between them shows a 5.8\% coding gain achieved by DCVC-FM over AVM. This may be because when the average BD-rate results were calculated for each sequence, the overlapped quality range varies if different anchor codec is employed.}. When VMAF is employed as the quality metric, neural video codecs, DCVC-DC and DCVC-FM demonstrate very similar performance as ECM, and all of them are better than AVM, with an evident (more than 8\%) coding gain in terms of BD-rate. Moreover, significant coding performance improvement has been achieved in each standard series, for example, ECM is nearly 20\% better than VTM, and a similar bitrate saving is also obtained by AVM over its predecessor, libaom.

\begin{figure*}[!t]
\centerline{\includegraphics[width=1\textwidth]{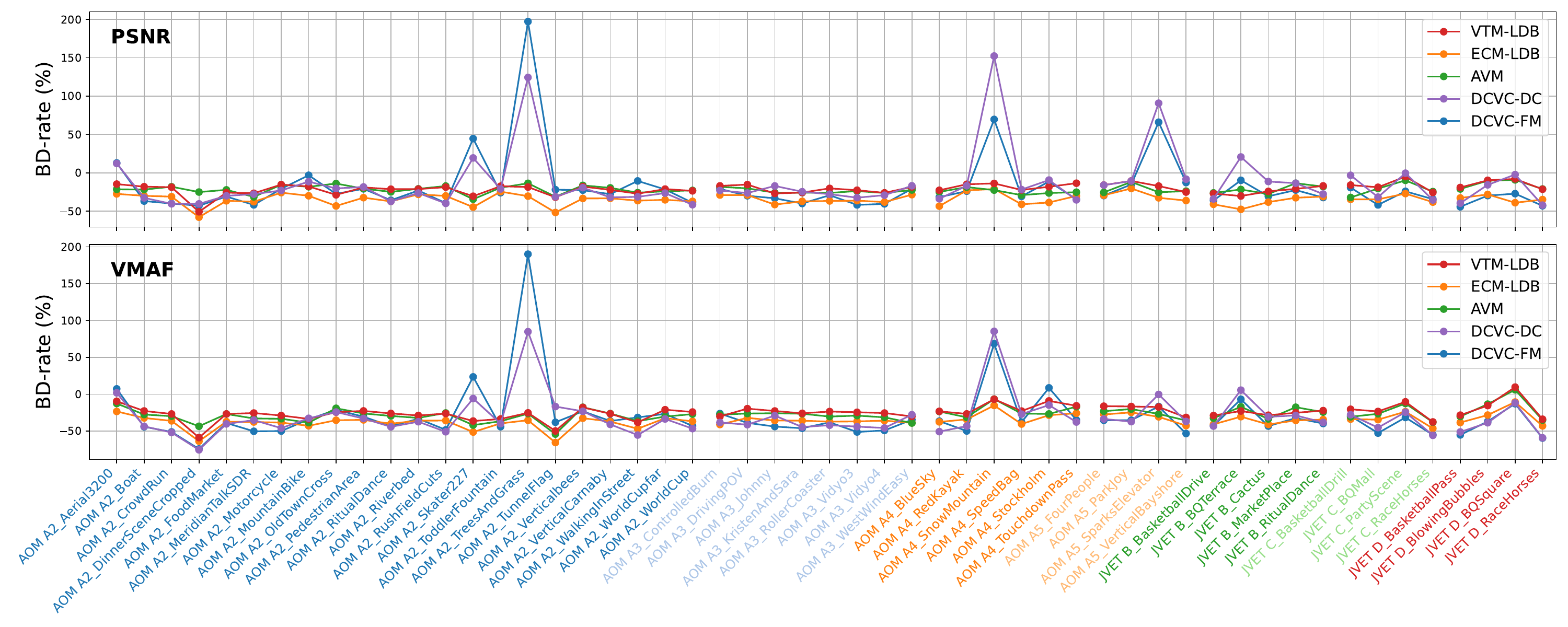}}
\vspace{-5pt}
\caption{Per sequence BD-rate results. Here libaom is used as the anchor.}
\label{fig: psnr per sequence}
\end{figure*}

\begin{figure*}[htbp]
    \centering    
    \subfigure{			
    \includegraphics[width=0.185\textwidth]{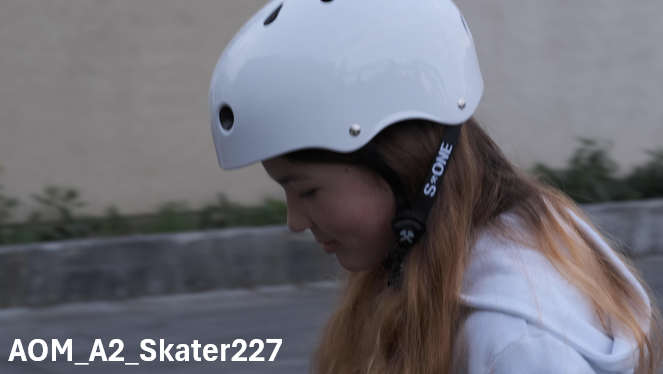}}
    \subfigure{			
    \includegraphics[width=0.185\textwidth]{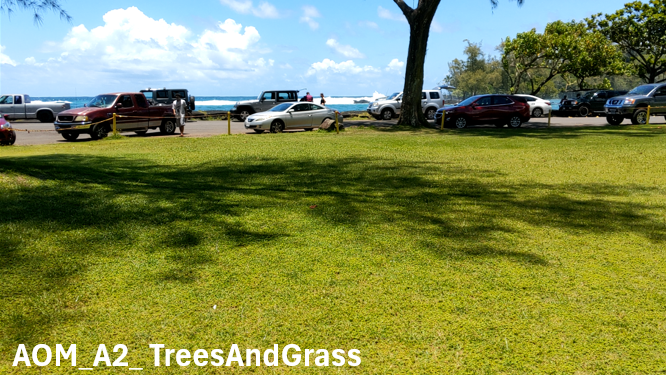}}	
    \subfigure{			
    \includegraphics[width=0.185\textwidth]{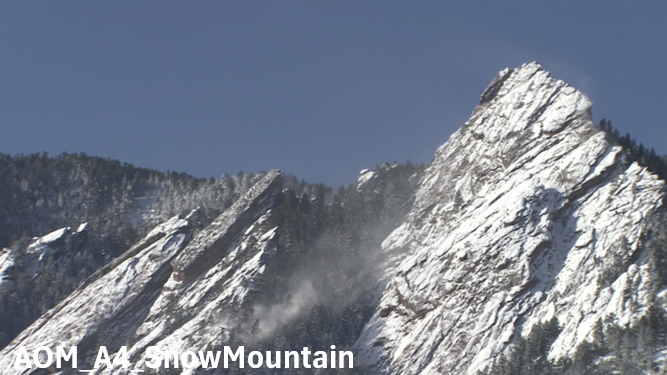}}	
    \subfigure{			
    \includegraphics[width=0.185\textwidth]{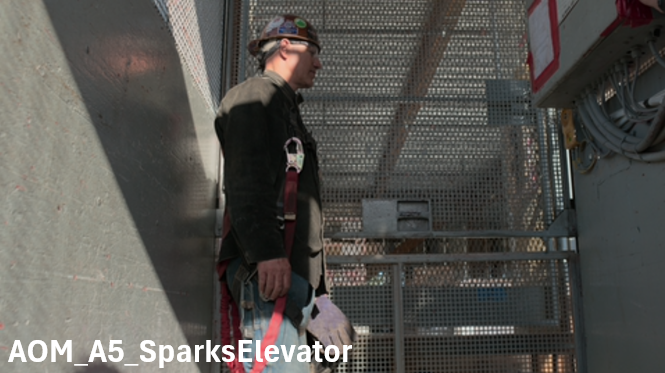}}
    \subfigure{			
    \includegraphics[width=0.185\textwidth]{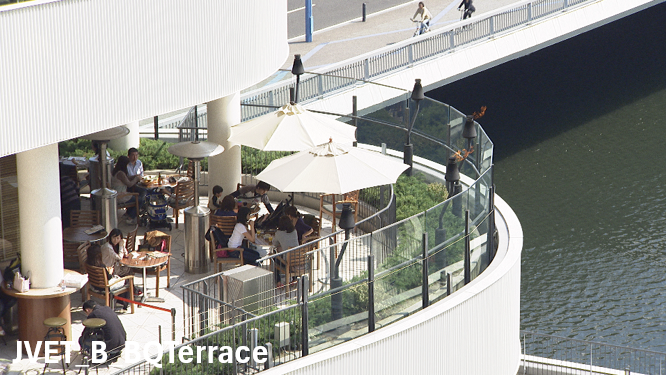}}	
    \caption{Example frames of the sequences on which two neural video codecs did not perform well.}		
    \label{Fig: dcvc scenes}								
\end{figure*}

We also illustrate coding performance by plotting the average rate quality curves for each codec, as shown in \autoref{Fig: rd-curve}. Here, in order to enable a meaningful comparison, we only use the results of 1080p sequences and selected the QP values of conventional video codecs to accommodate the limited range of two neural coding methods. We also excluded the LDP results for VTM and ECM, as they are outperformed by their LDB counterparts. Here, similar performance can be observed to the results in TABLE II - ECM is clearly the best performer, while VTM and two neural video codecs are superior to libaom.

\subsection{Per-resolution and Per-sequence Performance}
To further characterize the coding performance of the tested codecs across different resolutions and sequences, we summarize the BD-rate statistics based on PSNR and VMAF in \autoref{fig: psnr per sequence} for VTM-LDB, ECM-LDB, AVM, DCVC-FM, DCVC-DC, with libaom as the anchor. The results are also grouped by resolution, and shown in different colors. It can be observed that, in most cases, for the same codec,  greater savings are typically achieved on higher-resolution videos - this was also reported previously in \cite{perfComForHighRes}. Moreover, while the coding performance of conventional codecs is relatively consistent across different sequences, the two neural video codecs, DCVC-FM and DCVC-DC exhibit distinct characteristics, with significantly poorer performance on specific test content, such as \textit{Skater227} and \textit{SparksElevator} from AOM CTC Class A2 and A5, respectively. It is noted that these sequences exhibit large camera motions and further investigation is recommended in terms of model generalization improvement. Example frames of these sequences are shown in \autoref{Fig: dcvc scenes}.

\subsection{Coding Complexity}\label{coding complexity}

We report complexity comparisons of the six video codecs in TABLE II, based on relative encoding and decoding runtimes (using libaom as the anchor), based on HD content only. In this experiment, it should be noted that conventional codecs are executed based on CPU only, while neural codecs exploit additional GPU resources. We can observe that two neural codecs are much faster in encoding compared to all the conventional codecs tested, but their decoding speeds are slower than libaom, VTM and AVM. 

A radar chart plot is also shown in \autoref{fig:radar-chart} to illustrate the trade off between performance and run-time for VTM, ECM, AVM, DCVC-DC and DCVC-FM codecs. It is noted that none of these codecs offers a perfect balance between complexity and coding efficiency and that neural video coding methods have yet to achieve evident coding gains over conventional video codecs based on similar resource constraints.

\section{Conclusion}

This paper presents a thorough comparative study between six state-of-the-art conventional and neural video codecs in a Low-Delay configuration. The experiment is based on the well-established common test conditions used in MPEG JVET and AOM. Results show that JVET ECM delivers the best overall coding performance, with a 11\% coding gain over DCVC-FM and 16.1\% over AVM, but is associated with a relatively high encoding and decoding complexity. All the tested video codecs fail to provide a perfect trade off between complexity and performance, which emphasizes the urgency to develop high-performance and low-complexity video coding algorithms for low-delay applications. We also observed inconsistent performance in the case of the two learning-based video codecs on specific test content, highlighting the need for further investigation.

\section*{Acknowledgements}
The authors appreciate the funding from the China Scholarship Council, University of Bristol, Visionular Inc. and the UKRI MyWorld Strength in Places Programme (SIPF00006/1).

\small
\bibliographystyle{ieeetr}
\bibliography{egbib}

\end{document}